\pdfoutput=1
\documentclass[11pt,twoside,a4paper,cmspaper,final,collab]{cms-tdr}
\def\svnVersion{21f59b7}\def\svnDate{2025/02/18}\def\cmsCernNoTag{CERN-EP-2024-206}\def\cmsCernDate{\today}\def\cmsMessage{Published in the Journal of High Energy Physics as \href{http://dx.doi.org/10.1007/JHEP02(2025)050}{\doi{10.1007/JHEP02(2025)050}.}}
\begin{document}\cmsNoteHeader{SUS-23-008}

\newcommand{\recoil}{\ensuremath{U}\xspace}
\newcommand{\vecrecoil}{\ensuremath{\vec{U}}\xspace}
\newcommand{\ma}{\ensuremath{m_{\textrm{a}}}\xspace}
\newcommand{\mA}{\ensuremath{m_{\textrm{A}}}\xspace}
\newcommand{\mh}{\ensuremath{m_{\PSh}}\xspace}
\newcommand{\mH}{\ensuremath{m_{\PH}}\xspace}
\newcommand{\mHpm}{\ensuremath{m_{\PSHpm}}\xspace}
\newcommand{\mchi}{\ensuremath{m_{\PGc}}\xspace}
\newcommand{\WW}{\ensuremath{\PW\PW}\xspace}
\newcommand{\WZ}{\ensuremath{\PW\PZ}\xspace}
\newcommand{\ZZ}{\ensuremath{\PZ\PZ}\xspace}
\newcommand{\Zmumu}{\ensuremath{\PZ\to \PGmm\PGmp}\xspace}
\newcommand{\Zee}{\ensuremath{\PZ\to \Pem\Pep}\xspace}
\newcommand{\Zvv}{\ensuremath{\PZ\to\PGn\PGn}\xspace}
\newcommand{\Wlv}{\ensuremath{\PW\to\Pell\PGn}\xspace}
\newcommand{\Wmn}{\ensuremath{\PW\to \PGm\PGn}\xspace}
\newcommand{\Wen}{\ensuremath{\PW\to \Pe\PGn}\xspace}
\newcommand{\Wmunu}{\ensuremath{\PW(\PGm\PGn)}\xspace}
\newcommand{\Wenu}{\ensuremath{\PW(\Pe\PGne)}\xspace}
\newcommand{\Zll}{\ensuremath{\PZ(\Pellp\Pellm)}\xspace}
\newcommand{\Zmumujets}{\ensuremath{\PZ(\PGm\PGm)+\text{jets}}\xspace}
\newcommand{\Zeejets}{\ensuremath{\PZ(\Pe\Pe)+\text{jets}}\xspace}
\newcommand{\Znunujets}{\ensuremath{\PZ(\PGn\PGn)+\text{jets}}\xspace}
\newcommand{\Zlljets}{\ensuremath{\PZ(\Pell\Pell)+\text{jets}}\xspace}
\newcommand{\Zllgjet}{\ensuremath{\PZ(\Pell\Pell)+\gamma+\text{jet}}\xspace}
\newcommand{\Zjets}{\ensuremath{\PZ+\text{jets}}\xspace}
\newcommand{\Wjets}{\ensuremath{\PW+\text{jets}}\xspace}
\newcommand{\Wlnujets}{\ensuremath{\PW(\Pell\PGn)+\text{jets}}\xspace}
\newcommand{\Wmunujets}{\ensuremath{\PW(\PGm\PGn)+\text{jets}}\xspace}
\newcommand{\Wenujets}{\ensuremath{\PW(\Pe\PGn)+\text{jets}}\xspace}
\newcommand{\Wlnugjet}{\ensuremath{\PW(\Pell\PGn)+\gamma+\text{jet}}\xspace}
\newcommand{\ttbarjets}{\ensuremath{\ttbar+\text{jets}}\xspace}
\newcommand{\ttbarlnu}{\ensuremath{\ttbar(\Pell\PGn)}\xspace}
\newcommand{\coststar}{\ensuremath{\cos\Theta^{*}}\xspace}
\newcommand{\mdPhi}{\ensuremath{\min(\Delta\phi)}\xspace}
\newcommand{\dPhi}{\ensuremath{\Delta\phi}\xspace}
\newcommand{\mdPhiJM}{\ensuremath{\min(\Delta\phi(\text{jet, }\ptvecmiss))}\xspace}
\providecommand{\cmsLeft}{left\xspace}
\providecommand{\cmsRight}{right\xspace}
\providecommand{\PAGc}{{\HepAntiParticle{\PGc}{}{}}\xspace}
\providecommand{\PA}{{\HepParticle{A}{}{}}\xspace}

\cmsNoteHeader{SUS-23-008}

\title{Search for dark matter produced in association with a pair of bottom quarks in proton-proton collisions at \texorpdfstring{$\sqrt{s} = 13\TeV$}{sqrt(s) = 13 TeV}}

\date{\today}

\abstract{A search for dark matter (DM) particles produced in association with bottom quarks is presented. The analysis uses proton-proton collision data at a center-of-mass energy of $\sqrt{s}=13\TeV$, corresponding to an integrated luminosity of 138\fbinv. The search is performed in a final state with large missing transverse momentum and a pair of jets originating from bottom quarks. No significant excess of data is observed with respect to the standard model expectation. Results are interpreted in the context of a type-II two-Higgs-doublet model with an additional light pseudoscalar (2HDM+a). An upper limit is set on the mass of the lighter pseudoscalar, probing masses up to 260\GeV at 95\% confidence level. Sensitivity to the parameter space with the ratio of the vacuum expectation values of the two Higgs doublets, $\tan\beta$, greater than 15 is achieved, capitalizing on the enhancement of couplings between pseudoscalars and bottom quarks with high $\tan\beta$.}

\hypersetup{
	pdfauthor={CMS Collaboration},
	pdftitle={Search for dark matter produced in association with a pair of bottom quarks in proton-proton collisions at sqrt(s)=13 TeV},
	pdfsubject={CMS},
	pdfkeywords={CMS, dark matter, bottom quarks, 2HDM+a, pseudoscalar}}

\maketitle

\section{Introduction}\label{sec:intro}
So far, the evidence for the existence of dark matter (DM) has come from astrophysics. According to the lambda cold dark matter ($\Lambda$CDM) model of cosmology~\cite{Deruelle2018Aug}, the high-precision Planck satellite map of the cosmic microwave background~\cite{Planck:2018nkj} indicates that DM constitutes about 26\% of the total mass-energy of the known universe~\cite{Freese:2017idy}, however, the particle nature of DM has not been elucidated yet. If it is composed of subnuclear particles, then these could potentially be produced at the LHC. Such particles, whether elementary or composite, are weakly interacting with standard model (SM) particles, which means they might escape detection in collider experiments. Since these DM candidates are expected to be electrically neutral and stable, their presence could be inferred through an observed imbalance in the transverse momentum vectors of particles in the detector, known as missing transverse momentum (\ptvecmiss) with magnitude represented by \ptmiss.

Many beyond-the-SM (BSM) theories hypothesize the particle nature of DM~\cite{Feng:2010gw,Trimble:1987ee}. Based on astrophysical and cosmological indications and the observed relic density of the universe, weakly interacting massive particles (WIMPs)~\cite{Jungman:1995df}, with masses in the range from a few\GeV to a few\TeV, are a popular type of DM candidates that can be searched for at colliders. Such searches can be performed by looking for the pair production of WIMPs through a mediator that recoils against one or more visible SM particles. WIMPs can give a detector signature of the type $\PX+\ptmiss$~\cite{CMS:2024zqs}, where \PX is one or more SM particles, e.g. a heavy-quark pair, produced in association with DM.

In this paper, we perform a search for DM particles produced in association with a pair of bottom quarks ($\bbbar+\ptmiss$) using the proton-proton ($\Pp\Pp$) collision data collected with the CMS detector in 2016--2018, corresponding to an integrated luminosity of 138\fbinv. The search for DM produced in association with heavy-flavor quarks is complementary to other $\PX+\ptmiss$ searches and to searches for DM production in cascades in the context of supersymmetry (SUSY) at the LHC. The ATLAS and CMS Collaborations have performed DM searches using heavy-flavor jets as tags~\cite{CMS:2024zqs,ATLAS:2024kpy}, such as \Pb-jets from single top quarks ($\PQt/\PAQt$)~\cite{Sirunyan:2019gfm, 2011.09308, 2211.13138, 2402.16561}, \Pb-jets from top quark pairs ($\ttbar$)~\cite{Sirunyan:2018dub, 2211.05426}, or bottom quark pairs ($\bbbar$)~\cite{CMS:2017dcx, CMS:2019ybf,CMS:2018zjv, ATLAS:2017hoo, 2101.12527, 2108.13391}.

\section{Signal and background simulations}\label{sec:signalmodel}
\begin{figure}[htbp!]
	\centering
	\includegraphics{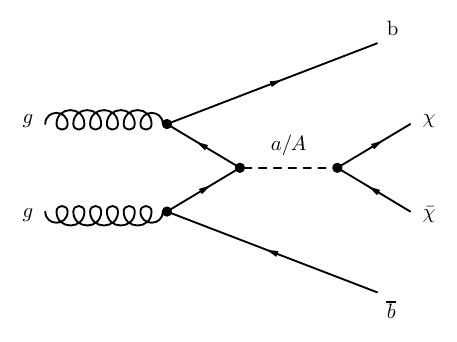}
	\includegraphics{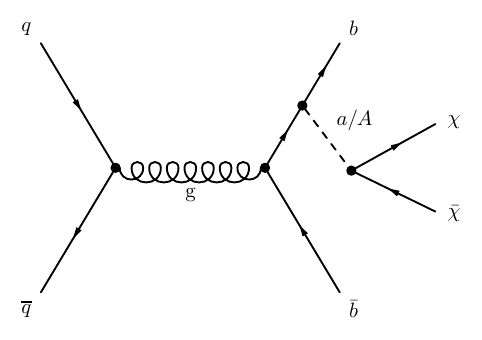}
	\caption{\label{fig:feynman_diagram} Leading Feynman diagrams for the $\bbbar+\PGc\PAGc$ process in the 2HDM+a model.}
\end{figure}

The model used for the interpretation of the results of this analysis is a type-II two-Higgs-doublet model with an additional pseudoscalar singlet interacting with DM particles, referred to as 2HDM+a~\cite{Abercrombie:2015wmb, Bauer:2017ota}. The model contains a scalar Higgs boson (\PSh, assumed to be the observed Higgs boson with mass $\mh = 125\GeV$), a pair of charged Higgs bosons (\PSHpm), a heavy scalar boson (\PH), a heavy pseudoscalar boson (\PSA), a light pseudoscalar boson (\Pa), and a Dirac fermion ($\PGc$), which is assumed to be the DM particle. Figure~\ref{fig:feynman_diagram} shows the leading Feynman diagrams for the production of the $\bbbar+\PGc\PAGc$ process within the 2HDM+a model.

The choice of the fixed and free parameters in this model follows the recommendations of Ref.~\cite{Abe:2018bpo}. The model contains the following parameters: the electroweak vacuum expectation value ($\nu\approx246\GeV$), the mass of the DM candidate $\mchi$, the masses of the Higgs bosons ($\mh, \mH, \mHpm$), the masses of the pseudoscalar bosons ($\ma, \mA$), the mixing angle between the two CP-even weak spin-0 eigenstates ($\alpha$), the mixing angle between the two CP-odd weak spin-0 eigenstates ($\theta$), the ratio of the vacuum expectation values of the two Higgs doublets ($\tan\beta$), and the quartic couplings ($\lambda_3, \lambda_{P1}, \lambda_{P2}$) of the scalar potential, which governs the interactions between the scalar doublets and the pseudoscalar.

In the decoupling limit, where $\sin(\beta-\alpha)=1$ and $\lambda_{3}=\lambda_{P_{1}}=\lambda_{P_{2}}=3$ ensure the boundedness of the Higgs potential, electroweak precision constraints give $m_{\PA}=m_{\PH}=m_{\PH^{\pm}}$, and the DM coupling $y_{\PGc}$ is set to 1. This results in five free parameters, namely,  $\tan\beta$, $\sin\theta$, $\mA$, $\ma$, and $\mchi$.

This choice is adopted in many 2HDM+a interpretations of heavy spin-0 resonances searches performed at the LHC~\cite{CMS:2018zjv, ATLAS:2017tlw,ATLAS:2017uhp,CMS:2018hir,ATLAS:2019wdu}. Furthermore, the motivation for using the 2HDM+a model is that the couplings of both the heavier pseudoscalar ($g_{\PSA\bbbar}$) and the lighter pseudoscalar ($g_{\Pa\bbbar}$) to bottom quarks are proportional to $\tan\beta$, resulting in increased interaction rates in the high-$\tan\beta$ region.

\begin{figure}[htb]
	\centering
	\includegraphics[width=0.49\textwidth]{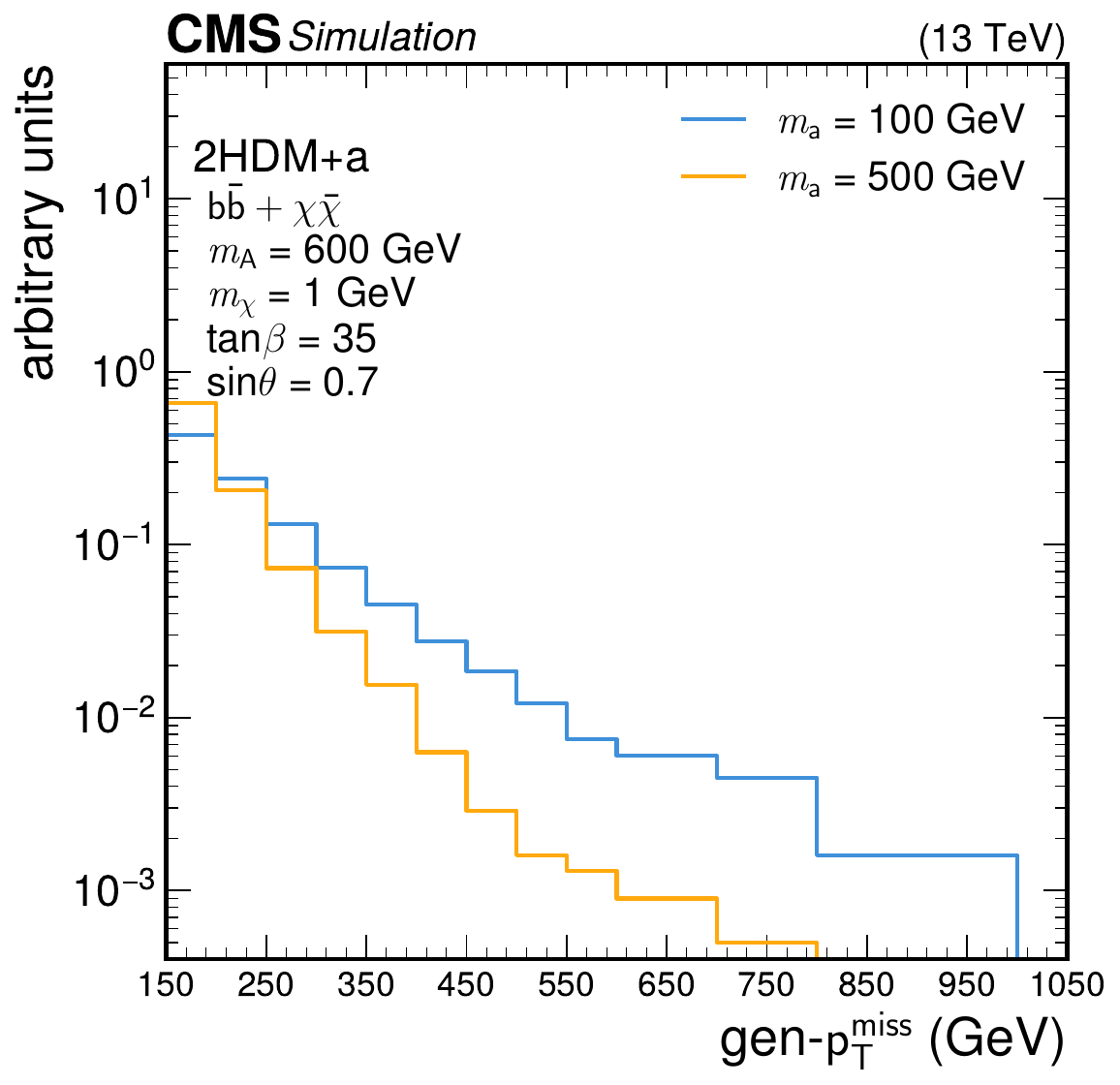}
	\includegraphics[width=0.49\textwidth]{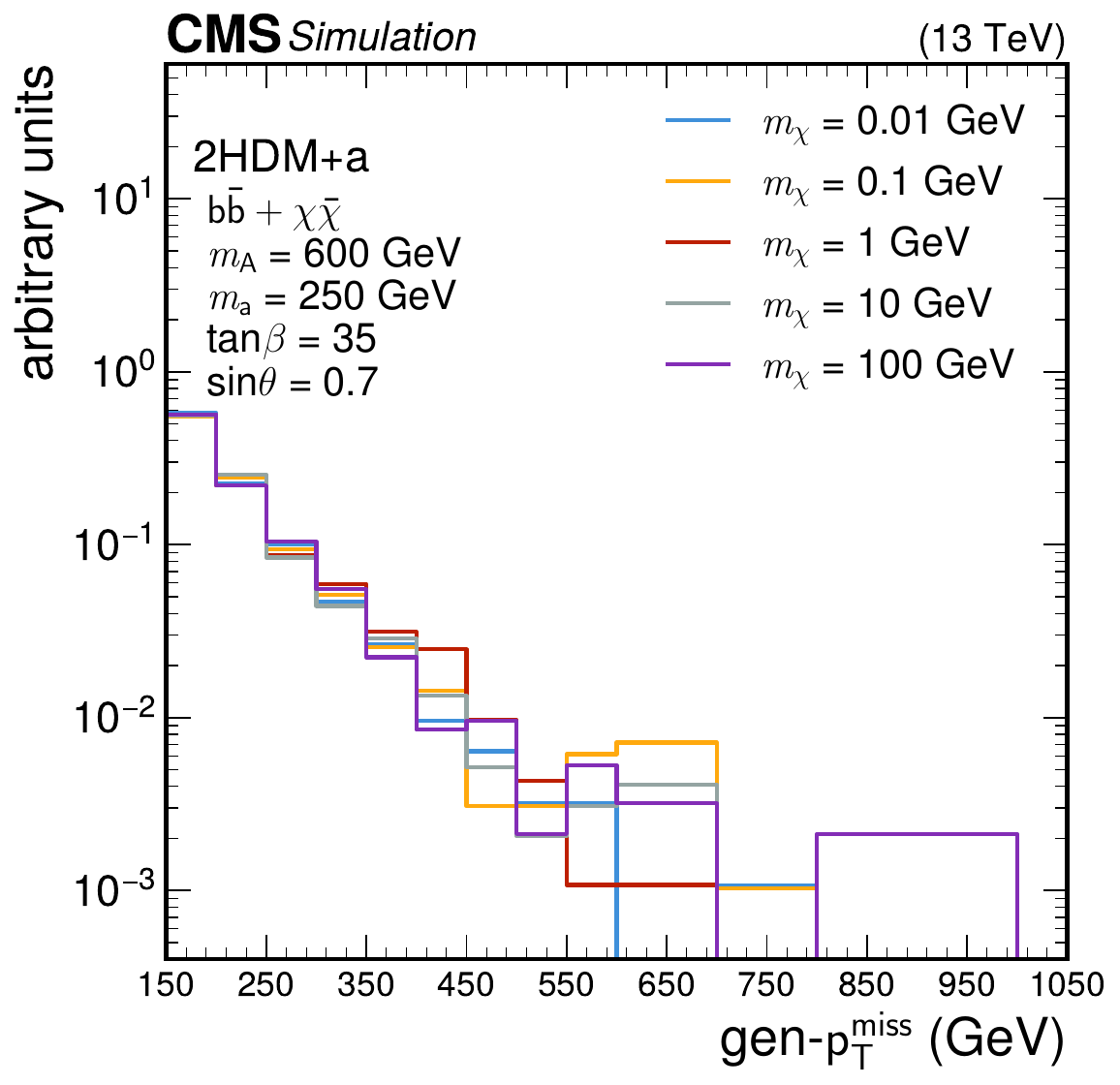}
	\caption{\label{fig:genMET_shape_ma}Normalized (to unity) shape of generator-level \ptmiss distribution for two illustrative lighter pseudoscalar masses \ma (\cmsLeft) and for five illustrative DM masses \mchi (\cmsRight).}
\end{figure}

The production cross section is dominantly driven by the mass of the lighter mediator, \Pa, as well as $\tan\beta$ and $\sin\theta$, with no strong dependence on the  mass of \PSA. Furthermore, as shown in Fig.~\ref{fig:genMET_shape_ma}, the distribution of \ptmiss strongly depends on $\ma$. On the other hand, $\tan\beta$, $\sin\theta$, and $\mchi$ marginally change the \ptmiss distribution, as also shown in Fig.~\ref{fig:genMET_shape_ma} for \mchi. Hence, signal events are generated with $\mA=600\GeV$, $\tan\beta=35$, $\sin\theta=0.7$, and $\mchi=1\GeV$ for different \ma values using \MGvATNLO v2.6.0~\cite{Alwall:2014hca} and the 5-flavor scheme at leading order (LO) in quantum chromodynamics (QCD) with no additional jets at the matrix element level.

A lower threshold of 150\GeV is set on \ptmiss at the generator level to constrain the kinematic phase space of simulated events to the region where the analysis is conducted. Various SM processes have a similar final state and higher production cross sections. The dominant background comes from the \Znunujets production ($\approx$60--70\% of the total background), when the \PZ boson is produced in association with one or two \PQb jets and decays to neutrinos. The \ttbar (lepton + jets) and \Wlnujets backgrounds are the next largest backgrounds, where a lepton coming from these processes is either lost or not reconstructed properly in the detector. Dileptonic and hadronic \ttbar decays, as well as single top quark and diboson (\WW, \WZ, \ZZ) production events, are smaller contributions to the total background.

The \Znunujets, \Zlljets, and \Wlnujets background processes ($\ell = \Pe,\PGm,\PGt$) are simulated using \MGvATNLO v2.4.2~\cite{Alwall:2014hca} at next-to-LO (NLO) in QCD. The \ttbar and single top quark production events are simulated using the \POWHEG v2 generator~\cite{Frixione:2007vw,Alioli:2010xd} at NLO in QCD. Diboson events are generated at LO in QCD using the \PYTHIA 8.212 generator~\cite{Sjostrand:2014zea}. The SM Higgs boson with a mass of 125\GeV produced in association with a vector boson is considered as a background and is simulated using the \POWHEG generator at NLO in QCD.
The NNPDF3.1 parton distribution functions (PDFs)~\cite{Ball:2014uwa} are used, depending on the QCD order of the generator for each physics process. Parton showering, fragmentation, and hadronization are simulated with \PYTHIA 8.230 (8.212 for 2016) using the CP5 (CUETP8M1 for 2016) underlying event tune~\cite{ue1,ue2}.

Interactions of the resulting final state particles with the CMS detector are simulated using the \GEANTfour software~\cite{Agostinelli:2002hh} and reconstructed using the same algorithms used for the data. Additional inelastic $\Pp\Pp$ interactions in the same or a neighboring bunch crossing (pileup) are included in the simulation. The pileup distribution is adjusted to match the corresponding distribution observed in data.

\section{The CMS detector}\label{sec:detector}

The central feature of the CMS apparatus is a superconducting solenoid of 6\unit{m} internal diameter, providing a magnetic field of 3.8\unit{T}. Within the solenoid volume are a silicon pixel and strip tracker, a lead tungstate crystal electromagnetic calorimeter (ECAL), and a brass and scintillator hadron calorimeter (HCAL), each composed of a barrel and two endcap sections. Forward calorimeters, made of steel and quartz fibers, extend the pseudorapidity ($\eta$) coverage provided by the barrel and endcap detectors. Muons are detected in gas-ionization chambers embedded in the steel flux-return yoke outside the solenoid.

Events of interest are selected using a two-tiered trigger system~\cite{CMS:2016ngn}. The first level, composed of custom hardware processors (L1), uses information from the calorimeters and muon detectors to select events at a rate of around 100\unit{kHz} in a fixed time interval of approximately 4\mus. The second level, known as the high-level trigger (HLT), consists of a farm of processors running a version of the full event reconstruction software optimized for fast processing, and reduces the event rate to around 1\unit{kHz} before data storage.

A more detailed description of the CMS detector, along with a definition of the coordinate system used and the relevant kinematic variables, can be found in Ref.~\cite{Chatrchyan:2008zzk}.

\section{Data samples and event reconstruction}\label{sec:datasamples_eventreconstruction}
This search uses data collected in $\Pp\Pp$ collisions between 2016 and 2018. In particular, physics data sets for which the events pass the \ptmiss and single-electron triggers at the HLT, and electron-photon and calo-\ptmiss triggers at L1 are used for final analysis.

A global ``particle-flow'' (PF) algorithm~\cite{CMS:2017yfk} aims to reconstruct all individual particles in an event, combining information provided by the inner tracker ECAL, HCAL, and muon detectors. The reconstructed particles are used to build \PGt leptons, jets, and missing transverse momentum~\cite{CMS:2018jrd,CMS:2016lmd,CMS:2019ctu}. The primary vertex is taken to be the vertex corresponding to the hardest scattering in the event, evaluated using tracking information alone, as described in Section~9.4.1 of Ref.~\cite{CMS-TDR-15-02}.

Jets are clustered using the anti-\kt jet clustering algorithm~\cite{Cacciari:2008gp} with a distance parameter ($\Delta R= \sqrt{\smash[b]{(\Delta \eta)^{2} + (\Delta \phi)^{2}}}$) of 0.4 (``AK4 jets''). Jet calibration \cite{CMS:2016lmd} is used to account for the energy deposits of neutral particles and full nonlinear hadronic response of the detector. Pileup effects from the charged hadrons are mitigated by removing them from pileup interactions, with a technique called ``charged-hadron subtraction'' (CHS)~\cite{CMS:2020ebo}. For mitigation of neutral PU particles, a new technique, pileup per particle identification (PUPPI)~\cite{Bertolini:2014bba} is used, which builds on the existing CHS algorithm. Furthermore, to ensure that the energy and momentum of the reconstructed jet matches that of the quark/gluon-initiated jet, jet energy corrections are applied to the reconstructed jets. In this search, jets with $\pt>30\GeV$ and $\abs{\eta}<2.5$ are considered. To remove any overlap with leptons, all jets within $\Delta R=0.4$ with leptons are removed.

Jets originating from bottom quark hadronization, \PQb jets, are identified with the  \textsc{DeepCSV} algorithm~\cite{BTV-16-002}, which uses a deep neural network (DNN) with 4 hidden layers and exploits the discriminating power of variables such as the position of the secondary vertex, the lifetime of the bottom quark, and the mass of b hadrons. Three working points are defined on the basis of the output of the DeepCSV algorithm. The loose working point corresponds to a 10\% misidentification ("mistag") rate with 84\% efficiency. The medium working point gives a 1\% mistag rate with 68\% efficiency, and the tight working point gives a 0.1\% mistag rate with 50\% efficiency.

The hadron-plus-strips algorithm is employed to reconstruct hadronically decaying $\Pgt$ leptons, denoted as \tauh~\cite{CMS:2018jrd}. Jets reconstructed using AK4 jet clustering are used as an input to the algorithm and $\tauh$ candidates are reconstructed on the basis of the number of charged hadrons and energy deposits in strips in the ECAL, which originate from the photon candidates and their electron-positron conversions. In order to minimize the misidentification rate of \tauh from quark and gluon jets, a convolutional discriminator based on a DNN called DeepTau~\cite{CMS:2022prd} is employed. Events with a \tauh candidate with $\pt > 20\GeV$ and  $\abs{\eta} < 2.3$ are vetoed.

Shower shape and isolation variables, as defined in Section 7.1 of Ref.~\cite{CMS:2020uim}, are used to identify electrons and photons. Isolated electrons and photons used in this analysis have $\pt > 10$ and 15\GeV, respectively, and $\abs{\eta} < 2.5$. Isolated muons have $\pt > 10\GeV$ and  $\abs{\eta} < 2.4$. Two parameters are considered while identifying the muons, namely the number of measurements in the tracker-muon system and the track reconstruction quality~\cite{CMS:2018rym}.

As mentioned in Section~\ref{sec:intro}, the negative vectorial \pt sum of all PF candidates is used to calculate \ptvecmiss with all jet energy corrections are applied. Events with anomalous \ptvecmiss can occur because of reconstruction failures, detector artifacts, or noncollision backgrounds~\cite{CMS:2019ctu}. Dedicated event filters, which can identify more than 90\% of such events, are used, with a mistagging rate of less than 0.1\%.

\section{Analysis strategy and event selection}
\begin{figure}[htbp!]
	\centering
	\includegraphics[width=0.49\textwidth]{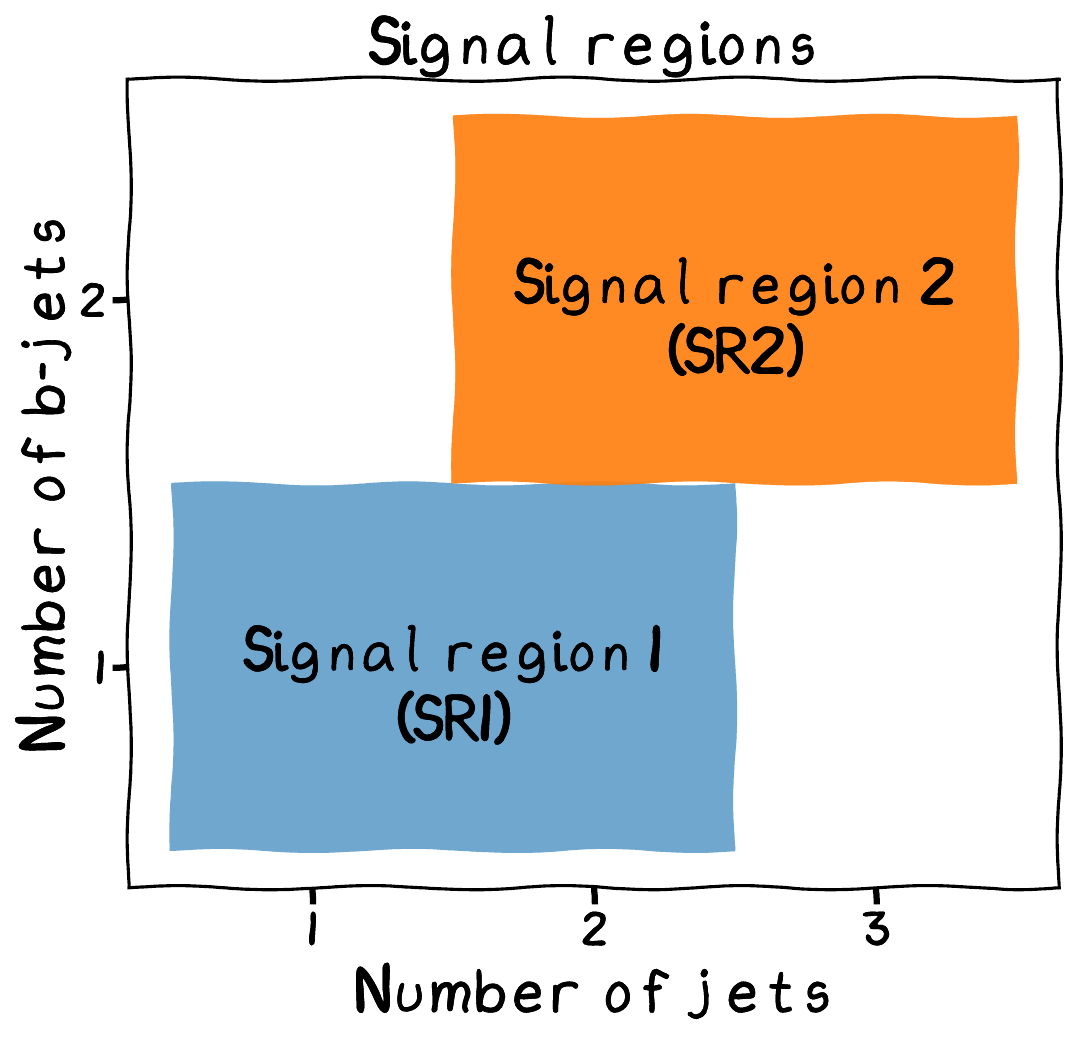}
	\includegraphics[width=0.49\textwidth]{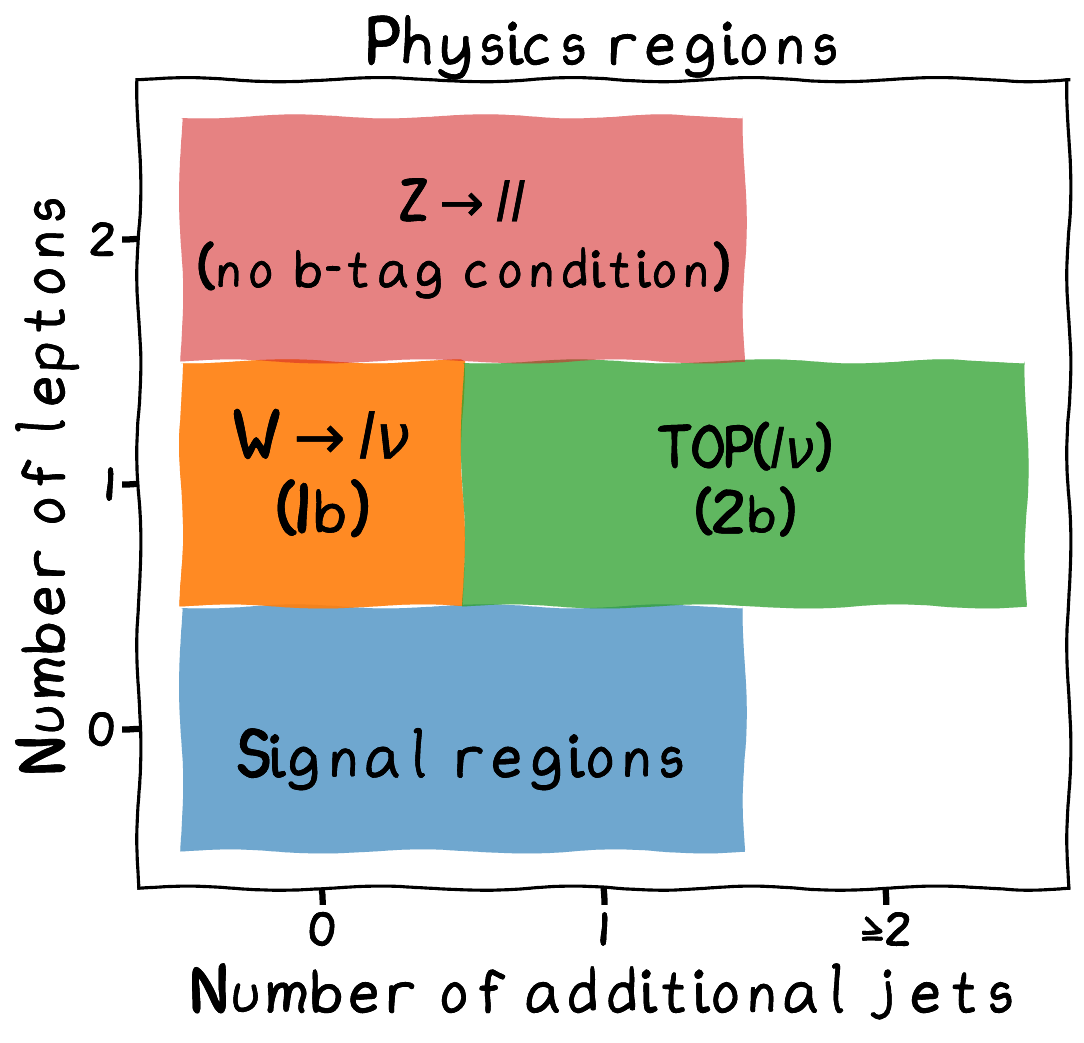}
	\caption{\label{fig:physics_regions_diagram} Definition of SR1 and SR2 in terms of jet multiplicity (\cmsLeft), and CR and SR definition in the lepton multiplicity and additional jet multiplicity plane (\cmsRight). The categories depend on the number and flavor of leptons, and the number of jets in an event.}
\end{figure}

The final states reconstructed in this analysis  consist of one or two \PQb-tagged jets and large \ptmiss, which could arise from DM particles. The data are split into two categories based on the number of \PQb-tagged jets: the 1b category contains events with only one \PQb-tagged jet and is composed of one signal region (SR), called SR1, and two control regions (CRs), while the 2b category contains events with two \PQb-tagged jets and is composed of one SR (SR2) and two CRs. Events that contain at least one well identified and isolated electron or muon with $\pt>10\GeV$ are vetoed from the SRs to reduce the background from the \ttbar and \Wlnujets processes. These regions are illustrated in Fig.~\ref{fig:physics_regions_diagram}.
The three dominant background sources \Znunujets, \ttbarlnu, and \Wlnujets are estimated by performing a simultaneous fit of the SRs and CRs, as detailed in Section \ref{sec:signal_extraction}.

In the CRs, \ptmiss is replaced by the absolute value of the transverse momentum recoil, $\vec{U}$, to mimic the variation observed in \ptmiss within the SRs because of the presence of leptons. The recoil is defined as:

\begin{equation}
	U = \abs{\vecrecoil} = \left|-\left(\ptvecmiss + \sum\vec{\pt}^{\text{lep}}\right)\right|
	\label{eq:def-recoil}
\end{equation}
In the 2b category, the \ptmiss distribution for signal events is similar to that of background events, such that there is little separation power for signal and background. The angular variable \coststar~\cite{Barr:2005dz}, defined as
\begin{equation}
	\coststar = \left|\tanh\left(\frac{\eta_{1} - \eta_2}{2}\right)\right|,
	\label{eq:costhetastar}
\end{equation}
where $\eta_1$ and $\eta_2$ are the pseudorapidities of the leading and subleading jets, gives an optimal discrimination between the signal and backgrounds in this category. In the longitudinally boosted frame, \coststar is the cosine of the polar angle between each jet and the beam axis, where the pseudorapidities of the jets are equal and opposite.

Events in the SRs and CRs are selected such that they are mutually exclusive. The events in the SRs and muon-based CRs are selected by a logical `OR' of \ptmiss triggers with different thresholds (90, 100, 110, 120, or 140 \GeV), whereas events in the electron-based CRs are selected by using a logical `OR' of single electron triggers with minimum threshold of 27, 32, or 35 \GeV~\cite{CMS:2016ngn}.
Events are preselected using a set of requirements common to the SRs and CRs. In the CRs, the selection criteria incorporate \recoil to mimic the variation observed in \ptmiss within the SRs because of the presence of leptons.

The event selection procedure starts with the preselection, which requires events passing trigger thresholds, followed by an offline selection of \ptmiss or \recoil $>250\GeV$. Events must include at least one jet, with the leading jet having $\pt > 100\GeV$, as optimized based on signal sensitivity. Events with a photon with $\pt>15\GeV$ and $\abs{\eta} < 2.5$ are rejected to suppress photon-induced backgrounds (\Zllgjet, \Wlnugjet).

QCD multijet events can mimic the signal signature when the jet momentum is mismeasured, giving rise to high \ptmiss. The QCD multijet background is reduced by requiring the minimum $\Delta\phi$ between any jet and \ptvecmiss $(\mdPhiJM)$ to be higher than 0.5 radians. This ensures that \ptvecmiss is not aligned with any jet, which is a common feature in events with mismeasured jets. The QCD multijet background is further reduced by requiring the relative difference in the value of \ptmiss measured with the PF algorithm  and with calorimeter only information to be less than 0.5. This quantity is defined as
\begin{equation*}
	\Delta\ptmiss(\text{PF - calorimeter}) = \begin{cases}
		\left|\ptmiss(\text{PF})/\ptmiss(\text{calorimeter}) - 1\right|,        & \text{in the SRs and } \\
		\left|\ptmiss(\text{PF})/\text{\recoil}(\text{calorimeter}) - 1\right|, & \text{ in the CRs.}
	\end{cases}
\end{equation*}
Table \ref{tab:preselection} summarizes the requirements to preselect the events for all three data-taking years.
A dedicated event selection for the SRs and CRs is applied on top of the preselection.

\begin{table}[ht!]
	\renewcommand{\arraystretch}{1.2}
	\topcaption{Preselection criteria applied to all the events entering the SRs and CRs.}
	\label{tab:preselection}
	\centering
	\begin{tabular}{lc}
		{Selection}                       & {Cut Value}                                          \\
		\hline
		Trigger                                  & \ptmiss or single electron                                  \\
		\ptmiss (\recoil in CR)                  & $>$250\GeV                                                 \\
		Jets                                     & multiplicity $ \geq 1$, $\pt >30 \GeV$ , $\abs{\eta} < 2.5$ \\
		Leading jet                              & $\pt >100\GeV$                                              \\
		Overlap removal                          & no overlap with $\Pe$, $\Pgm$, $\tauh$, $\PGg$              \\
		$\PGg$                                   & veto photon with  $\pt >15\GeV$ and $\abs{\eta} < 2.5$      \\
		\mdPhiJM                                 & $>$0.5                                                      \\
		$\Delta\ptmiss(\text{PF - calorimeter})$ & $<$0.5                                                     \\
	\end{tabular}
\end{table}

Events in SR2 have two jets tagged as originating from bottom quarks (using the medium working point) and large \ptmiss. Furthermore, up to one additional jet with $\pt>30\GeV$, not \PQb-tagged, is allowed in events entering this region.
SR1 is foreseen to catch signal events where one of the \PQb jets is lost because it is outside of the detector acceptance or is not tagged correctly, requiring exactly one \PQb-tagged jet and up to one additional jet. This category helps to recover some lost signal efficiency at the cost of a relatively higher background.

\section{Background estimation}\label{sec:background_estimation}
To estimate the \Znunujets background, two CRs, targeting the \Zee and \Zmumu processes, are defined. In these CRs, out of exactly two leptons in the event, the leading lepton must have $\pt>30\GeV$, the invariant mass of the dilepton candidate must be between 70 and 110\GeV, and there is no \PQb tagging requirement on the jets. The CRs dedicated to the \Wlnujets and \ttbar backgrounds are built for SR1 and SR2, respectively. They require a single lepton (electron or muon) with $\pt>30\GeV$ and a transverse mass $\text{m}_{\text{T}} =\sqrt{2\ptmiss\pt^{l}(1-\cos[\Delta\phi(\ptvecmiss,\vec{\pt}^{l})])}$ less than 160\GeV, together with the same \PQb tagging requirements as in the SRs. In addition, events in the \ttbar CR have at least 1 additional non \PQb-tagged jet, while, for the \Wlnujets CR, exactly one \PQb-tagged jet is required and \ptmiss must be greater than 100\GeV.

\begin{figure}[htbp!]
	\centering
	\includegraphics[width=0.49\textwidth]{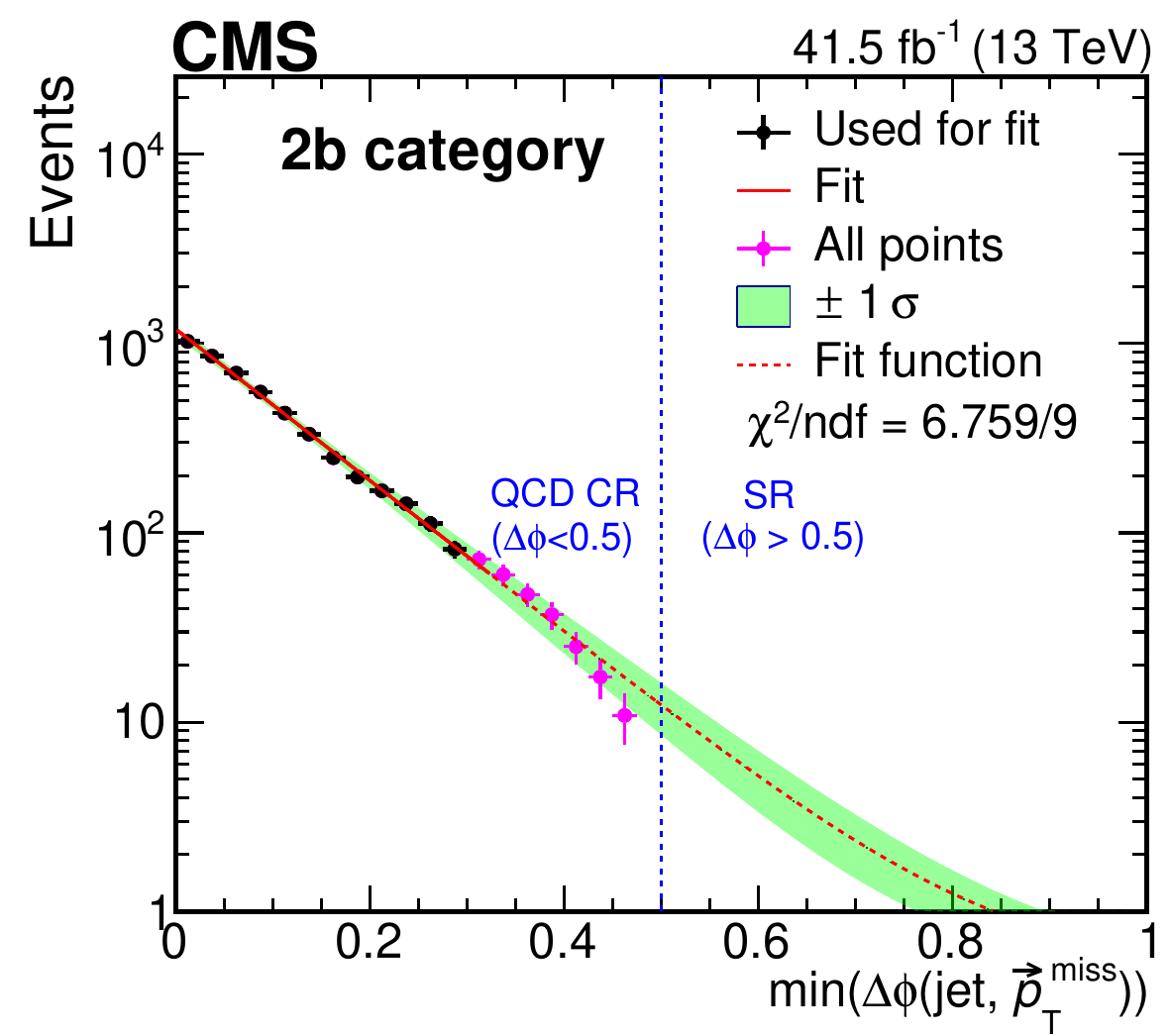}
	\caption{\label{fig:qcd_extrapolation2017} QCD background contribution in the QCD CR (black and pink dots) in the 2b category using 2017 data.}
\end{figure}

To estimate the QCD multijet background, a QCD CR is defined using the same selection as the SR except for requiring $\mdPhiJM<0.5$. The QCD background in the CR is estimated as the difference between data and other simulated backgrounds, and is plotted against the \mdPhi threshold, i.e., for each entry of \mdPhi on the $x$ axis, the $y$ axis corresponds to the number of events passing the \mdPhi  requirement. This distribution is plotted for each bin of \ptmiss or \coststar and then fitted with a function that describes the behavior of the distribution. The QCD contribution exhibits an exponential decrease with \mdPhi, as shown for one example in Fig.~\ref{fig:qcd_extrapolation2017}.
The QCD background is fitted with an exponential in the \mdPhi range between 0 to 0.3, and the resulting function is validated in the \mdPhi range between 0.3 to 0.5. The QCD contribution in the SR is calculated as the integral of the function for $\mdPhiJM>0.5$.

\section{Signal extraction}\label{sec:signal_extraction}
The fit for the 1b category is performed using \ptmiss in SR1 and \recoil in corresponding CRs as observables, therefore, both SR1 and the corresponding CRs enter simultaneously in a binned profile likelihood fit to estimate the signal. Each bin of \ptmiss in SR1 is tied to the corresponding bin of \recoil in the CR via a scale factor, treating the systematic uncertainties as nuisance parameters in the fit. A variable binning is used for \ptmiss and \recoil with bin edges 250, 300, 400, 550, and 1000 \GeV.

In the 2b category, the variable \coststar is used with bins of 0.25 width from 0.0 to 1.0, and a similar procedure is used to constrain the backgrounds as shown in Figure~\ref{fig:stack_run2_2b} (Z and \ttbar) using equivalent bins in the CRs. For each category, the electron and muon CRs are combined to form a single CR for each background estimated from data.

Furthermore, the fit to the data determines the background post-fit yields in each bin, as shown in the lower panels of Figure~\ref{fig:stack_run2_1b} and \ref{fig:stack_run2_2b}, displaying the ratio of the yields in data over the prediction, before and after the fit. The pre-fit yield values are determined by running the selection on simulated events to determine selection efficiency, and normalizing them to the integrated luminosity times their respective theoretical cross sections. The post-fit values are obtained as the pre-fit yields adjusted by the post-fit values of the nuisance parameters and scale factors. The lower panels of Figure~\ref{fig:stack_run2_1b} and \ref{fig:stack_run2_2b} show that the post-fit background distributions agree significantly better with the data in the control regions, confirming that the method is working.

\section{Systematic uncertainties}\label{sec:systematics_uncertainties}
The background and signal predictions are affected by systematic uncertainties. Only the systematic uncertainties that have an impact of at least half a percent are taken into account in the signal extraction fit. Uncertainties that do not impact the shape of the distribution, or are uniformly distributed, are treated as normalization uncertainties, whereas those that exhibit non-uniform variations are treated as shape and normalization uncertainties (except for scale and pdf uncertainties, where only shape has been taken into account).

Shape-based uncertainties in the \PQb tagging efficiency are calculated separately for true \PQb jets (having \PQb flavor at the particle level) and incorrectly \PQb-tagged jets for each event.

The uncertainty in the integrated luminosity amounts to 1.2--2.5\%, depending on the data-taking year~\cite{CMS-LUM-17-003,CMS-PAS-LUM-17-004,CMS-PAS-LUM-18-002}, while the overall uncertainty for the 2016--2018 period is 1.6\%.

The \ptmiss-based trigger efficiency scale factors are measured in a phase space with single-muon and dimuon events, and the difference in the scale factors is considered as the systematic uncertainty. A shape-based systematic uncertainty of the order of 1\% is applied to all the simulated samples in the SRs and muon CRs. The electron trigger efficiency scale factors are calculated using a tag-and-probe method~\cite{CMS:2010svw} and a shape-based uncertainty is applied to all the simulated samples in the electron CRs. The overall effect is of the order of 2--3\% on the normalization.

Shape uncertainties arising from jet energy corrections are applied to the simulations, separating different sources of uncertainty. The average effect is of the order of 2--3\%.

An uncertainty in the total inelastic cross section of 4.6\%~\cite{CMS:2018mlc} is used as a shape-based uncertainty to account for the effects of pileup modeling.

For the single top quark, diboson, and SM Higgs boson processes, a normalization uncertainty of 10\%, 20\%, and 20\%, respectively, is used~\cite{CMS:2014fut,CMS:2016ogx,CMS:2021icx}.

The systematic uncertainty in the cross section due to the renormalization and factorization scales is taken into account for signal and all backgrounds, using the envelope of the distributions treated as the shape uncertainty. The uncertainty related to the modeling of PDFs is estimated using the replicas provided in the corresponding PDF sets for the signal and  different backgrounds~\cite{Butterworth2016Jan,Dulat2016Feb,Harland-Lang2015May,Ball:2014uwa}.

For the QCD multijet background, the total uncertainty comes from the combination of the uncertainty obtained from the 95\% confidence intervals of the fitted extrapolation function and a flat  uncertainty of 20\% that covers the statistical uncertainty. The total effect on the QCD background estimation is about 50\%. The maximum variations of the sources of systematic uncertainties across individual bins and processes, along with their respective categories, is detailed in Table~\ref{tab:systematics}.

\begin{table}[htbp!]
	\centering
	\caption{Maximum variations from systematic uncertainties, affecting the shape and normalization ($\ln$N) of the signal and background predictions. Only the systematic sources with an impact greater than 0.5\% are listed.\\ $^{*}$refers to corresponding individual processes }
	\resizebox{0.6\textwidth}{!}{
		\begin{tabular}{lcc}
			{Systematics}                     & {Type} & {Size} \\
			\hline
			Pileup reweighting                             & shape$+\ln$N         & 3\%           \\
			\PQb tag efficiency                            & shape$+\ln$N         & 8\%           \\
			Incorrectly \PQb-tagged jets                   & shape$+\ln$N         & 7\%           \\
			Electron identification                        & shape$+\ln$N         & 7\%           \\
			Electron reconstruction efficiency             & shape$+\ln$N         & 5\%           \\
			Muon identification                            & shape$+\ln$N         & $<$1\%        \\
			Muon isolation                                 & shape$+\ln$N         & $<$1\%        \\
			Electron trigger                               & shape$+\ln$N         & 1\%           \\
			\ptmiss trigger                                & shape$+\ln$N         & 1\%           \\
			Jet energy correction                          & shape$+\ln$N         & 3\%           \\
			L1 prefiring                                   & shape$+\ln$N         & 0.5\%         \\
			Muon prefiring                                 & $\ln$N               & 0.5\%         \\
			$\tauh$ veto                                   & $\ln$N               & 6\%           \\
			Integrated luminosity                          & $\ln$N               & 2.5\%         \\
			$^{*}$Cross section diboson                    & $\ln$N               & 20\%          \\
			$^{*}$Cross section Drell--Yan                 & $\ln$N               & 5\%           \\
			$^{*}$Cross section Higgs boson                & $\ln$N               & 20\%          \\
			$^{*}$Cross section single top quark           & $\ln$N               & 10\%          \\
			$^{*}$QCD normalization                        & $\ln$N               & 50\%          \\
			Renormalization and factorization scales       & shape                & $<$42\%       \\
			PDF                                            & shape                & $<$50\%       \\
		\end{tabular}
	}
	\label{tab:systematics}
\end{table}

\section{Results}\label{sec:resutls}
\begin{figure}[htbp!]
	\centering
	\includegraphics[width=0.84\textwidth]{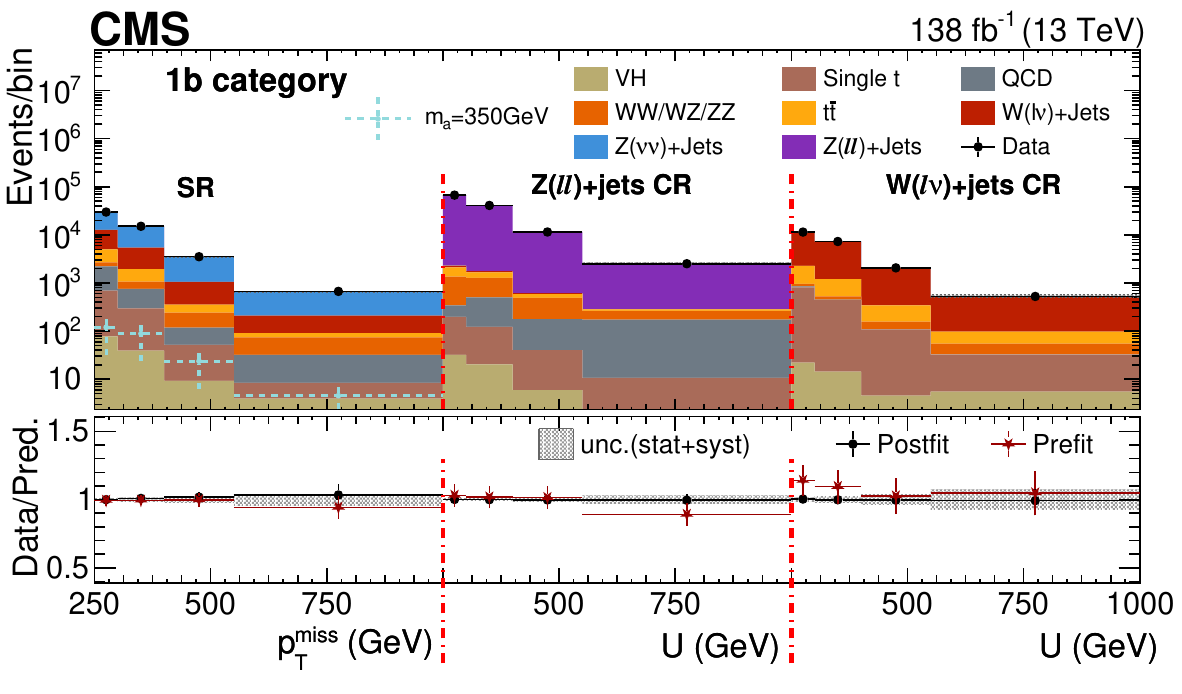}
	\caption{\label{fig:stack_run2_1b} Observed and predicted \ptmiss (\recoil) distributions in the 1b category. The red lines divide the plot into three parts: the first part shows the SR, the second part the \Zlljets CR, and the third part the \Wlnujets CR. The bottom plot shows the ratio of observed data to predicted distributions for both postfit and prefit predictions. The uncertainty band represents the combined systematic and statistical uncertainties in the predictions. Prefit points are shown with their combined systematic and statistical uncertainties.}
\end{figure}

\begin{figure}[htbp!]
	\centering
	\includegraphics[width=0.84\textwidth]{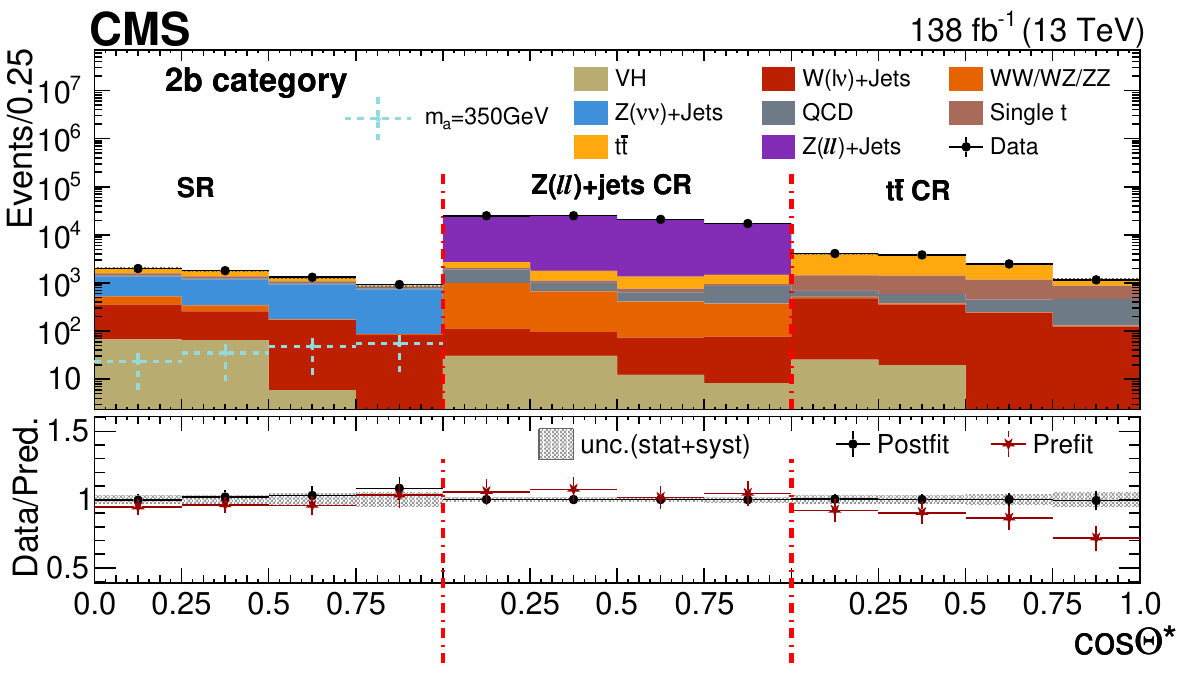}
	\caption{\label{fig:stack_run2_2b} Observed and predicted \coststar distributions in the 2b category. The red lines divide the plot into three parts: the first part shows the SR, the second part the \Znunujets CR, and the third part the \ttbar CR. The bottom plot shows the ratio of observed data to predicted distributions for both postfit and prefit predictions. The uncertainty band represents the combined systematic and statistical uncertainties in the predictions. Prefit points are shown with their combined systematic and statistical uncertainties.}
\end{figure}

Figures~\ref{fig:stack_run2_1b} and \ref{fig:stack_run2_2b} present the expected distributions of \ptmiss for SR1 and \recoil for the \Zlljets and \Wlnujets CRs in the 1b category, as well as \coststar for the SR and CRs in the 2b category. These distributions are shown both before (prefit) and after (postfit) the maximum likelihood fit, with the observed data superimposed. No excess with respect to the SM prediction is observed.

\begin{figure}[htbp!]
	\centering
	\includegraphics[width=0.49\textwidth]{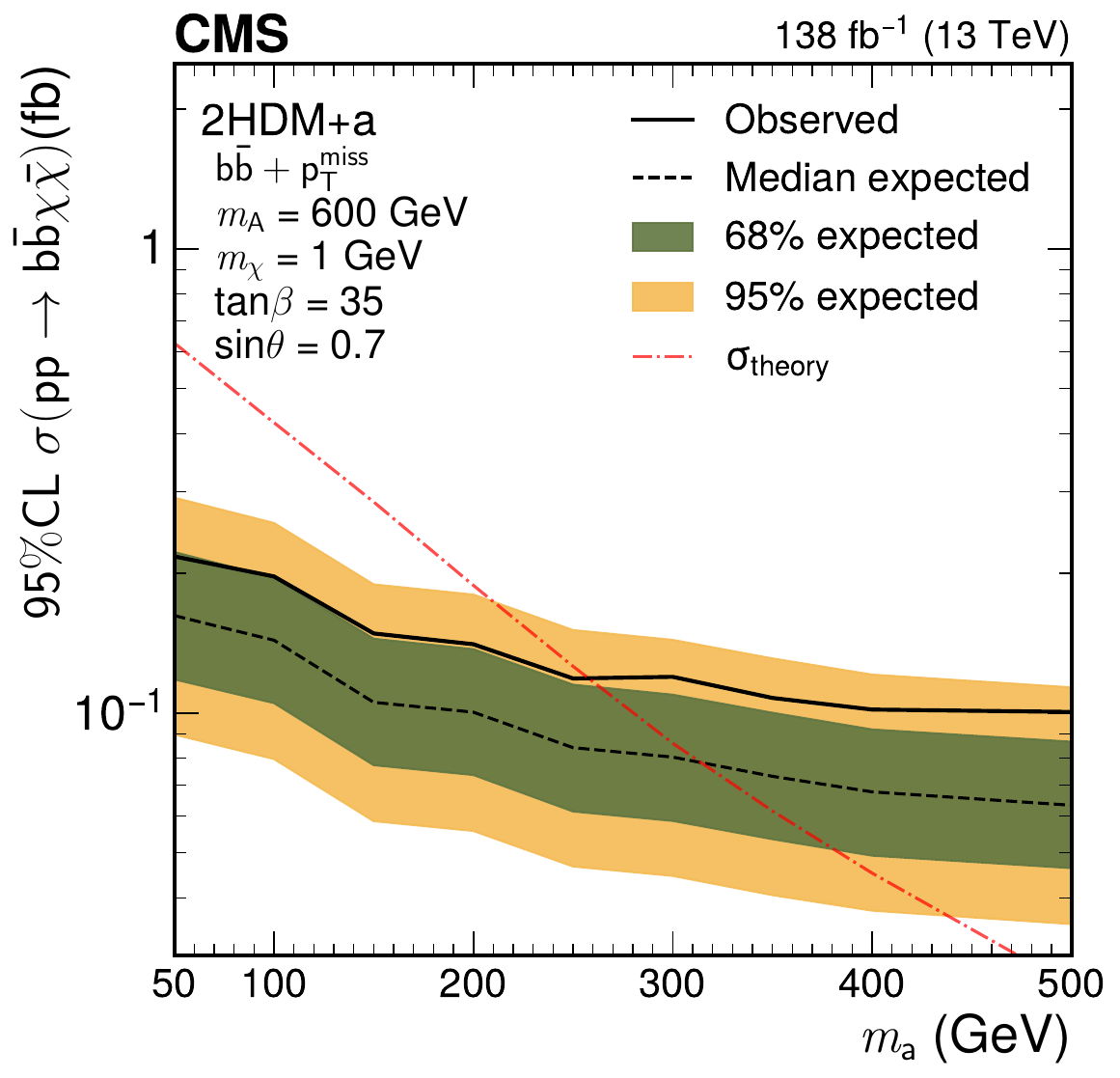}
	\caption{\label{fig:lim_run2_1d} The 95\% CL upper limit on the production cross-section of DM candidates in association with a pair of bottom quarks for $\mA=600\GeV$, $\sin\theta=0.7$, $\mchi=1\GeV$, and $\tan\beta=35$, for the combination of SR1 and SR2. The green and yellow bands show the $\pm 1$ and $\pm 2$ standard deviations from expected limits. The mass points below the red line are excluded.}
\end{figure}

Upper limits on the production crossection of DM candidates in association with two bottom quarks as predicted by the theoretical model, $\sigma/$, are extracted at the 95\% confidence level (CL) using a modified frequentist method~\cite{Heinemeyer:2013tqa, bib:CLS1, bib:CLS2} computed with an asymptotic approximation~\cite{bib:CLS3}. The combined limit is shown in Fig.~\ref{fig:lim_run2_1d}. The observed exclusion range for \ma is up to $260\GeV$, for  $\mA=600\GeV$, $\tan\beta=35$, $\sin\theta=0.7$, and $\mchi=1\GeV$.

\begin{figure}[htbp!]
	\centering
	\includegraphics[width=0.49\textwidth]{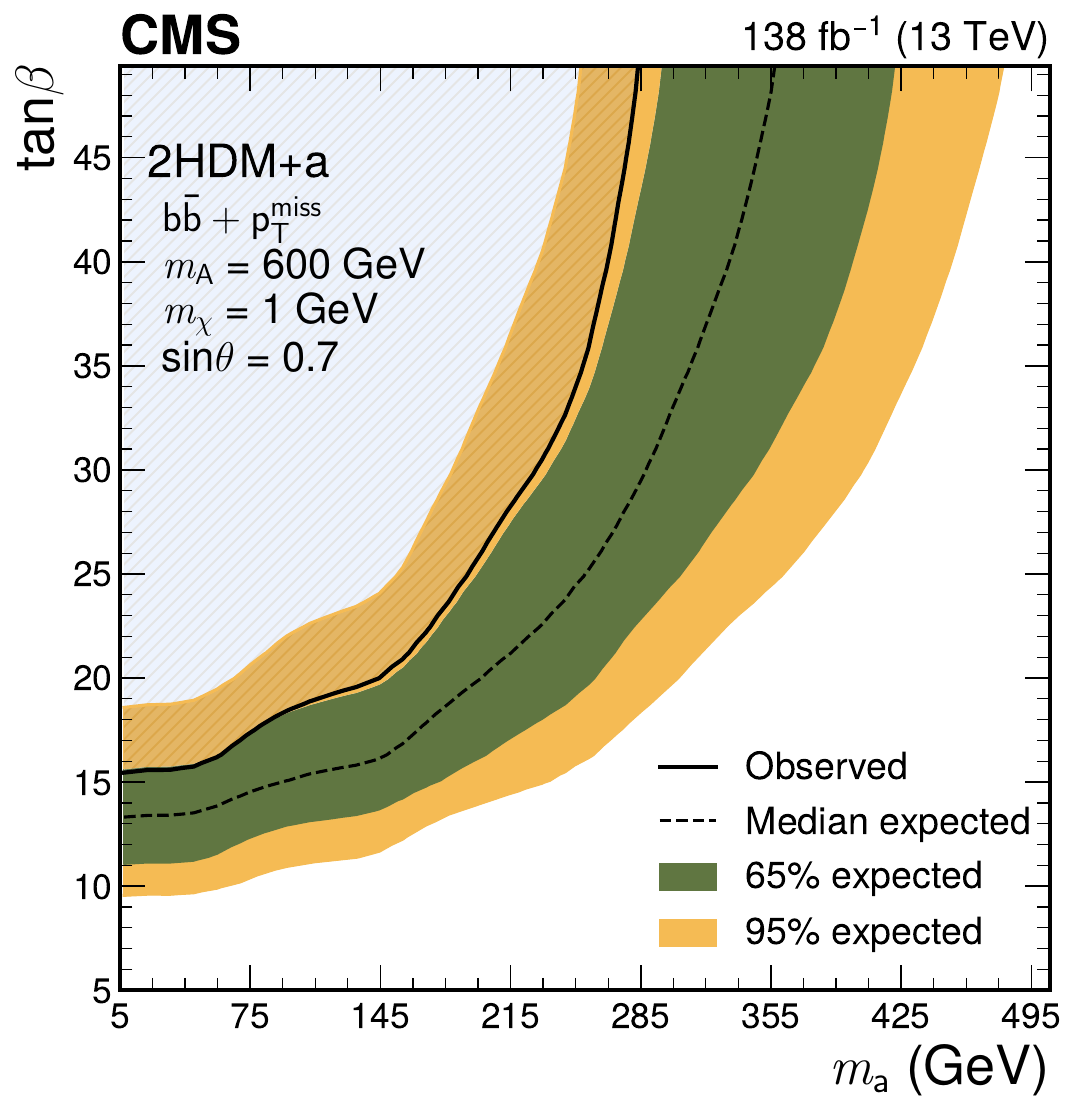}
	\includegraphics[width=0.49\textwidth]{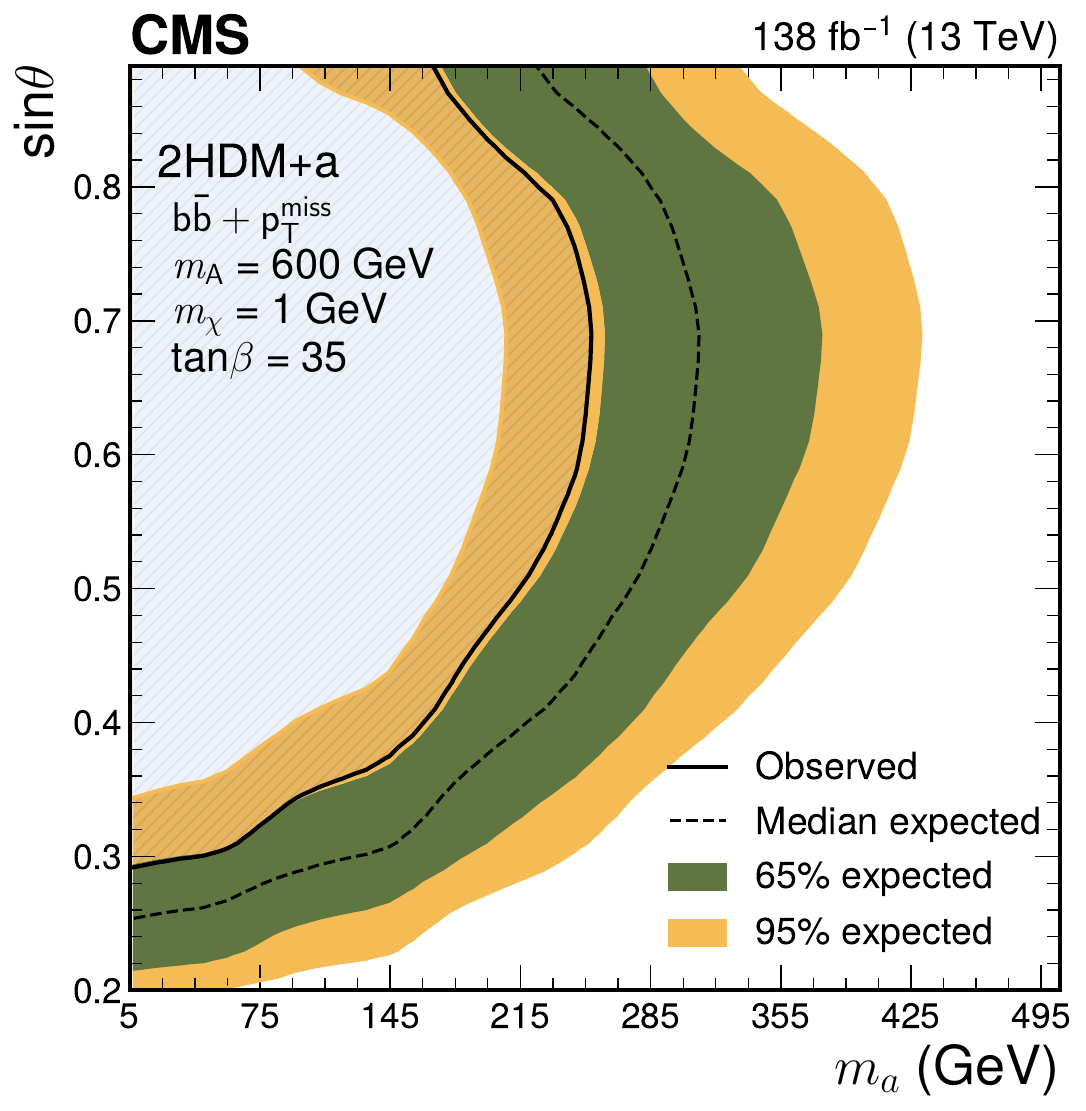}\\
	\includegraphics[width=0.49\textwidth]{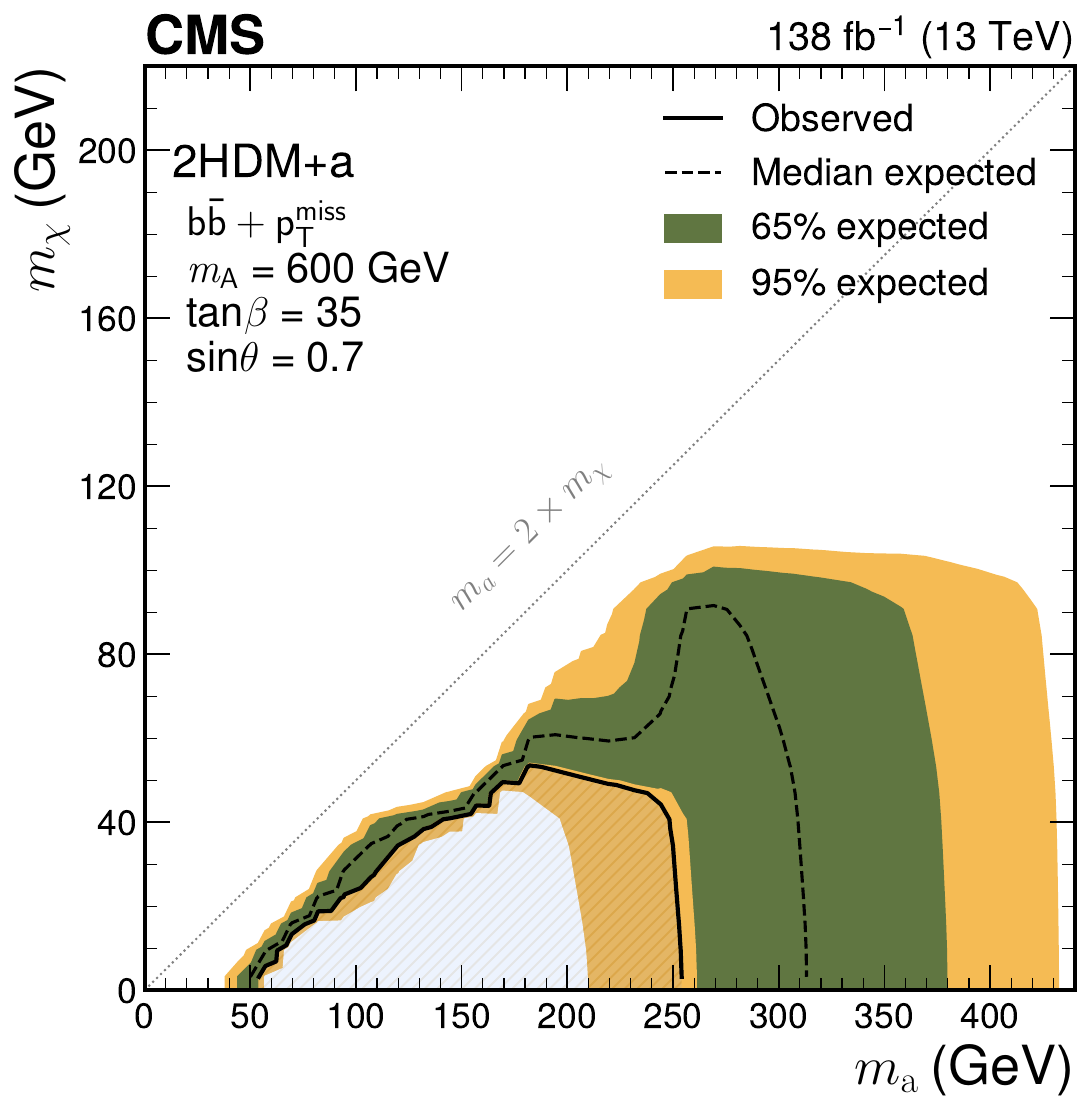}
	\caption{\label{fig:lim_run2_2d} Observed and expected upper limits at 95\% CL on the signal strength in the \ma -- $\tan\beta$ (upper left), \ma -- $\sin\theta$ (upper right), and \ma -- \mchi (lower) planes. The shaded area bounded by solid black line is excluded, with, the dotted grey line ($\ma = 2\mchi$) separating on-shell and off-shell productions for the lower plot.}
\end{figure}

The two-dimensional limit on signal strength in the plane of \ma-$\tan\beta$, using $\mA=600\GeV$, $\sin\theta=0.7$ and $\mchi=1\GeV$, is shown in  Fig.~\ref{fig:lim_run2_2d} (upper left). The two-dimensional limit in the plane of \ma-$\sin\theta$, using $\mA=600\GeV$, $\tan\beta=35$ and $\mchi=1\GeV$, is shown in  Fig.~\ref{fig:lim_run2_2d} (upper right).  The two-dimensional limit in the plane of \ma--\mchi, using $\tan\beta=35$, $\sin\theta=0.7$, and $\mA=600\GeV$ is shown in Fig.~\ref{fig:lim_run2_2d} (lower).

Tabulated results are provided in the HEPData record for this analysis~\cite{hepdata}.

\section{Summary}
A search for dark matter produced in association with a pair of bottom quarks is performed using data collected by the CMS detector in 2016--2018, corresponding to an integrated luminosity of 138\fbinv. The analysis searches for a possible signal by using two independent categories with different multiplicities of jets reconstructed as originating from a bottom quark. The results are interpreted in the framework of a simplified model, namely a type-II two-Higgs-doublet model with an additional pseudoscalar singlet (2HDM+a). It is the first search at the LHC to probe dark matter (DM) produced in association with two nonresonant bottom quarks in this model. This search performs best in the high-$\tan\beta$ phase space, where the signal production cross section is enhanced. This provides a complementary search to constrain the 2HDM+a model parameter phase space in the region where other DM searches are less sensitive. The lighter pseudoscalar mass below 260\GeV is excluded at 95\% confidence level for $\tan\beta=35$, $\sin\theta=0.7$, and $\mA=600\GeV$.

\begin{acknowledgments}
	We congratulate our colleagues in the CERN accelerator departments for the excellent performance of the LHC and thank the technical and administrative staffs at CERN and at other CMS institutes for their contributions to the success of the CMS effort. In addition, we gratefully acknowledge the computing centers and personnel of the Worldwide LHC Computing Grid and other centers for delivering so effectively the computing infrastructure essential to our analyses. Finally, we acknowledge the enduring support for the construction and operation of the LHC, the CMS detector, and the supporting computing infrastructure provided by the following funding agencies: SC (Armenia), BMBWF and FWF (Austria); FNRS and FWO (Belgium); CNPq, CAPES, FAPERJ, FAPERGS, and FAPESP (Brazil); MES and BNSF (Bulgaria); CERN; CAS, MoST, and NSFC (China); MINCIENCIAS (Colombia); MSES and CSF (Croatia); RIF (Cyprus); SENESCYT (Ecuador); ERC PRG, RVTT3 and MoER TK202 (Estonia); Academy of Finland, MEC, and HIP (Finland); CEA and CNRS/IN2P3 (France); SRNSF (Georgia); BMBF, DFG, and HGF (Germany); GSRI (Greece); NKFIH (Hungary); DAE and DST (India); IPM (Iran); SFI (Ireland); INFN (Italy); MSIP and NRF (Republic of Korea); MES (Latvia); LMTLT (Lithuania); MOE and UM (Malaysia); BUAP, CINVESTAV, CONACYT, LNS, SEP, and UASLP-FAI (Mexico); MOS (Montenegro); MBIE (New Zealand); PAEC (Pakistan); MES and NSC (Poland); FCT (Portugal); MESTD (Serbia); MCIN/AEI and PCTI (Spain); MOSTR (Sri Lanka); Swiss Funding Agencies (Switzerland); MST (Taipei); MHESI and NSTDA (Thailand); TUBITAK and TENMAK (Turkey); NASU (Ukraine); STFC (United Kingdom); DOE and NSF (USA).

	\hyphenation{Rachada-pisek} Individuals have received support from the Marie-Curie program and the European Research Council and Horizon 2020 Grant, contract Nos.\ 675440, 724704, 752730, 758316, 765710, 824093, 101115353, 101002207, and COST Action CA16108 (European Union); the Leventis Foundation; the Alfred P.\ Sloan Foundation; the Alexander von Humboldt Foundation; the Science Committee, project no. 22rl-037 (Armenia); the Belgian Federal Science Policy Office; the Fonds pour la Formation \`a la Recherche dans l'Industrie et dans l'Agriculture (FRIA-Belgium); the F.R.S.-FNRS and FWO (Belgium) under the ``Excellence of Science -- EOS" -- be.h project n.\ 30820817; the Beijing Municipal Science \& Technology Commission, No. Z191100007219010 and Fundamental Research Funds for the Central Universities (China); the Ministry of Education, Youth and Sports (MEYS) of the Czech Republic; the Shota Rustaveli National Science Foundation, grant FR-22-985 (Georgia); the Deutsche Forschungsgemeinschaft (DFG), among others, under Germany's Excellence Strategy -- EXC 2121 ``Quantum Universe" -- 390833306, and under project number 400140256 - GRK2497; the Hellenic Foundation for Research and Innovation (HFRI), Project Number 2288 (Greece); the Hungarian Academy of Sciences, the New National Excellence Program - \'UNKP, the NKFIH research grants K 131991, K 133046, K 138136, K 143460, K 143477, K 146913, K 146914, K 147048, 2020-2.2.1-ED-2021-00181, and TKP2021-NKTA-64 (Hungary); the Council of Science and Industrial Research, India; ICSC -- National Research Center for High Performance Computing, Big Data and Quantum Computing and FAIR -- Future Artificial Intelligence Research, funded by the NextGenerationEU program (Italy); the Latvian Council of Science; the Ministry of Education and Science, project no. 2022/WK/14, and the National Science Center, contracts Opus 2021/41/B/ST2/01369 and 2021/43/B/ST2/01552 (Poland); the Funda\c{c}\~ao para a Ci\^encia e a Tecnologia, grant CEECIND/01334/2018 (Portugal); the National Priorities Research Program by Qatar National Research Fund; MCIN/AEI/10.13039/501100011033, ERDF ``a way of making Europe", and the Programa Estatal de Fomento de la Investigaci{\'o}n Cient{\'i}fica y T{\'e}cnica de Excelencia Mar\'{\i}a de Maeztu, grant MDM-2017-0765 and Programa Severo Ochoa del Principado de Asturias (Spain); the Chulalongkorn Academic into Its 2nd Century Project Advancement Project, and the National Science, Research and Innovation Fund via the Program Management Unit for Human Resources \& Institutional Development, Research and Innovation, grant B39G670016 (Thailand); the Kavli Foundation; the Nvidia Corporation; the SuperMicro Corporation; the Welch Foundation, contract C-1845; and the Weston Havens Foundation (USA).
\end{acknowledgments}

\bibliography{auto_generated}
\cleardoublepage \appendix\section{The CMS Collaboration \label{app:collab}}\begin{sloppypar}\hyphenpenalty=5000\widowpenalty=500\clubpenalty=5000
\cmsinstitute{Yerevan Physics Institute, Yerevan, Armenia}
{\tolerance=6000
A.~Hayrapetyan, A.~Tumasyan\cmsAuthorMark{1}\cmsorcid{0009-0000-0684-6742}
\par}
\cmsinstitute{Institut f\"{u}r Hochenergiephysik, Vienna, Austria}
{\tolerance=6000
W.~Adam\cmsorcid{0000-0001-9099-4341}, J.W.~Andrejkovic, T.~Bergauer\cmsorcid{0000-0002-5786-0293}, S.~Chatterjee\cmsorcid{0000-0003-2660-0349}, K.~Damanakis\cmsorcid{0000-0001-5389-2872}, M.~Dragicevic\cmsorcid{0000-0003-1967-6783}, P.S.~Hussain\cmsorcid{0000-0002-4825-5278}, M.~Jeitler\cmsAuthorMark{2}\cmsorcid{0000-0002-5141-9560}, N.~Krammer\cmsorcid{0000-0002-0548-0985}, A.~Li\cmsorcid{0000-0002-4547-116X}, D.~Liko\cmsorcid{0000-0002-3380-473X}, I.~Mikulec\cmsorcid{0000-0003-0385-2746}, J.~Schieck\cmsAuthorMark{2}\cmsorcid{0000-0002-1058-8093}, R.~Sch\"{o}fbeck\cmsorcid{0000-0002-2332-8784}, D.~Schwarz\cmsorcid{0000-0002-3821-7331}, M.~Sonawane\cmsorcid{0000-0003-0510-7010}, W.~Waltenberger\cmsorcid{0000-0002-6215-7228}, C.-E.~Wulz\cmsAuthorMark{2}\cmsorcid{0000-0001-9226-5812}
% ...
\par}
\vskip\cmsinstskip
\dag:~Deceased\\
$^{1}$Also at Yerevan State University, Yerevan, Armenia\\
$^{2}$Also at TU Wien, Vienna, Austria\\
% ...
\end{sloppypar}
\end{document}